\documentclass[twocolumn,secnumarabic,amssymb, nobibnotes, aps, prd, superscriptaddress,showpacs]{revtex4}

\usepackage[dvips]{graphicx}

\begin{document}

\title{Ro-vibrational analysis of the XUV photodissociation of HeH$^+$ ions}

\author{J. Loreau}
\email{jloreau@gmail.com}
\affiliation{Institute for Theoretical Atomic, Molecular and Optical Physics, Harvard-Smithsonian Center for Astrophysics, Cambridge, Massachusetts 02138, USA}
\author{J. Lecointre}
\email{julien.lecointre@uclouvain.be}
\affiliation{Institute of Condensed Matter and Nanosciences, Universit\'e catholique de Louvain, 1348 Louvain-la-Neuve, Belgium}
\author{X. Urbain}
\email{xavier.urbain@uclouvain.be}
\affiliation{Institute of Condensed Matter and Nanosciences, Universit\'e catholique de Louvain, 1348 Louvain-la-Neuve, Belgium}
\author{N. Vaeck}
\email{nvaeck@ulb.ac.be}
\affiliation{Laboratoire de Chimie Quantique et Photophysique, Universit\'e Libre de Bruxelles, CP160/09, 1050 Bruxelles, Belgium}

%\date{\today}%

\begin{abstract}

We investigate the dynamics of the photodissociation of the hydrohelium cation HeH$^+$ by XUV radiation with the aim to establish a detailed comparison with a recent experimental work carried out at the FLASH free electron laser using both vibrationally hot and cold ions.
As shown in previous theoretical works, the comparison is hindered by the fact that the experimental ro-vibrational distribution of the ions is unknown. We determine this distribution using a dissociative charge transfer set-up and the same source conditions as in the FLASH experiment.
Using a non-adiabatic time-dependent wave packet method, we calculate the partial photodissociation cross sections for the $n=1-3$ coupled electronic states of HeH$^+$. We find a good agreement with the experiment for the total cross section into the He + H dissociative channel. By performing an adiabatic calculation involving the $n=4$ states, we then show that the experimental observation of the importance of the electronic states with $n>3$ cannot be well explained theoretically, especially for cold ($v=0$) ions. We also calculate the relative contributions to the cross section of the $\Sigma$ and $\Pi$ states. The agreement with the experiment is excellent for the He$^+$ + H channel, but only qualitative for the He + H$^+$ channel. We discuss the factors that could explain the remaining discrepancies between theory and experiment.

\end{abstract}


\maketitle
%\tableofcontents

\section{Introduction}

The hydrohelium cation HeH$^+$ has always attracted the attention of theorists. With only two electrons and two nuclei, it is one of the simplest heteronuclear molecular ions and it has been widely used as a benchmark for theoretical methods. In addition, being composed of the two most abundant species in the universe, it has many astrophysical applications. HeH$^+$ is thought to be the first molecular ion to form in the early universe, by radiative association of He and H$^+$ \cite{Lepp2002}. It has been predicted to be abundant in various astrophysical objects such as planetary nebulae \cite{Dabrowski1978}, helium-rich white dwarfs \cite{Harris2004} or metal-poor stars \cite{Engel2005}, although to this day it still eludes astrophysical detection. Several studies have been devoted to the search for HeH$^+$ in the planetary nebula NGC 7027 (see \cite{Liu1997} for a review), but it has appeared that its detection would be difficult.
The abundance of HeH$^+$ in astrophysical environments is determined by the balance between creation and destruction mechanisms. In the presence of a central star, such as in the case of planetary nebulae, the dominant destruction process of HeH$^+$ is its photodissociation by energetic UV photons. It can occur trough two dissociative pathways:
\begin{eqnarray}\label{photodissociation}
\mathrm{HeH}^+ (X ^1\Sigma^+) + h\nu \longrightarrow (\mathrm{HeH}^+)^*
&& \longrightarrow \mathrm{He}(1snl\ ^1L) + \mathrm{H}^+ \nonumber \\
&& \longrightarrow \mathrm{He}^+(1s) + \mathrm{H}(nl) \nonumber \\
&&
\end{eqnarray}
As a consequence, several theoretical studies of the photodissociation process were realized. The first theoretical estimate of the photodissociation cross section is due to Saha {\it et al.} \cite{Saha1978}, who considered the transition to the first excited $A\ ^1\Sigma^+$ state. The rate constant for the photodissociation process into the $A$ state was first calculated by Roberge and Dalgarno \cite{Roberge1982} using more accurate molecular data, while a calculation including the $B\  ^1\Sigma^+$ state and the lowest $^1\Pi$ state was done by Basu and Barua \cite{Basu1984}.

In 2007, the first experimental data on the photodissociation of HeH$^+$ in its ground $X\ ^1\Sigma^+$ state were obtained using the free-electron laser (FEL) FLASH in Hamburg \cite{Pedersen2007}, operating in the XUV at a wavelength of 32 nm (corresponding to an energy of 38.74 eV). One of the main interests of free electron lasers is the possibility to operate at wavelengths below 100 nm, which allows to probe excited electronic states of molecules. The experiment made at FLASH consists in a crossed beam photodissociation of HeH$^+$ with three-dimensional reaction imaging. The events leading to neutral helium fragments were detected, and the kinetic energy release (KER) of these fragments was analyzed, along with the angular orientation of the dissociating molecule with respect to the photon polarization. The photodissociation cross section was also determined.
This experiment on the HeH$^+$ system revealed the importance of $n\geq3$ states in the photodissociation, as well as the dominance of photodissociation perpendicular to the laser polarization and therefore the major role of the $\Pi$ states during dissociation. This demonstrated the necessity of taking into account the perpendicular orientation generally neglected or incompletely considered in previous theoretical models. Following the publication of these results, two theoretical studies investigating the role of excited states in the photodissociation process were realized. Dumitriu and Saenz \cite{Dumitriu2009} calculated the cross section using a time-independent approach and neglecting the non-adiabatic couplings. Sodoga {\it et al.} \cite{Sodoga2009} computed the cross section using a time-dependent approach and showed the importance of these couplings in the photodissociation process, but considered only states up to a principal quantum number $n=3$. Despite these extensive calculations, an agreement with the experiment was difficult to obtain, one of the main reasons being that the experimental ro-vibrational distribution of the HeH$^+$ ions was not known. While both theoretical works concluded that the cross section was not significantly affected by rotational excitation, the effect of vibrational excitation was found to be substantial. The most important unknown quantity is therefore the vibrational distribution of the ions in the experimental source.

In this paper, we build on the work of Sodoga {\it et al.} with the aim to achieve a more detailed comparison with the experiment. By reproducing the ion source conditions, we determine the ro-vibrational distribution of the experimental ions in the FLASH experiment, and we use it to compare the theoretical and experimental cross sections.
In the meantime, Pedersen {\it et al.} published new results on the photodissociation of HeH$^+$ \cite{Pedersen2010}. In this new study, the authors produced a source of cold ions by trapping the ions for about 100 ms before studying the photodissociation process. After such a long storage time, it can effectively be considered that all the ions are in the $v=0$ state. Moreover, the authors were able to detect events leading to the dissociation into He$^+$ + H. For hot (vibrationally excited) and cold ($v=0$) ions, the authors measured the branching ratio between the two dissociation pathways in (\ref{photodissociation}) as well as the relative contribution of the $^1\Sigma^+$ and $^1\Pi$ states to the process.
By investigating theoretically the photodissociation process for both hot and cold ions, we will show that there exists significant discrepancies between the theoretical and experimental results. In particular, we will argue that states with $n>3$ should not play a major role in the photodissociation process at the experimental energy. We will attempt to find the various factors that could explain these conflicting results.

This paper is organized as follows: in section \ref{theory}, we quickly summarize the theoretical method used to treat the photodissociation process. In section \ref{vib_distrib}, we describe the experimental method which allowed the determination of the ro-vibrational distribution. In section \ref{results}, we compare our results with the experimental data, and we discuss the possible reasons for the discrepancy between theory and experiment.

\section{Theoretical framework}\label{theory}

We will only recall here the principal steps in the theoretical description of the photodissociation process. The reader is referred to references \cite{Sodoga2009} and \cite{Loreau2010a} for a complete account of the theoretical methods used in this work. Our approach is based on the separation of the electronic and nuclear motions. The first step is to solve the electronic Schr\"odinger equation using {\it ab initio} methods, giving access to the potential energy curves (PEC) and non-adiabatic coupling matrix elements. The photodissociation cross section is then obtained using a time-dependent approach, which consists in the propagation of wave packets on the coupled electronic states.

\subsection{Electronic structure of HeH$^+$}\label{ab_initio}

The ground state of the HeH$^+$ molecular ion is a $^1\Sigma^+$ state. In the dipole approximation, the photodissociation can only occur through excited $^1\Sigma^+$ (parallel orientation of the field with respect with the molecular axis) and $^1\Pi$ states (perpendicular orientation).

\begin{figure}
%\hspace{-1cm}
\includegraphics[angle=-90,width=.5\textwidth]{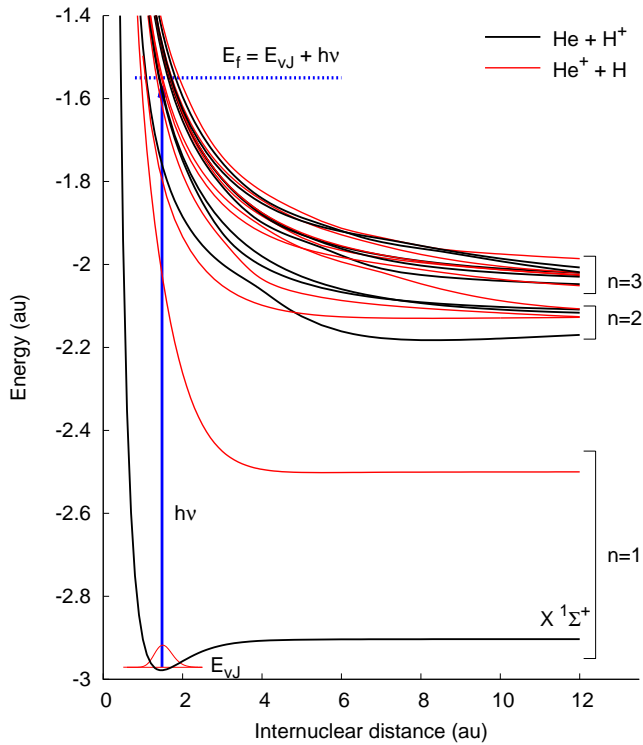}
\caption{Illustration of the photodissociation of HeH$^+$. The ion is initially in a ro-vibrational level of the $X\ ^1\Sigma^+$ ground state characterized by an energy $E_{vJ}$ and a wave function $\psi_{vJ} (R)$ (in this example, $v=0, J=0$). The photoabsorption is denoted by a vertical arrow, and the dotted line corresponds to the experimental energy ($h\nu = 38.74$ eV). We show the adiabatic potential energy curves of all the $n=1-3$ $^1\Sigma^+$ and $^1\Pi$ states through which the photodissociation can occur.
}
\label{fig_PEC}
\end{figure}

We considered here all the $n=1-3$ $^1\Sigma^+$ and $^1\Pi$ states of the HeH$^+$ molecular ion. The potential energy curves for these states have been calculated using the {\it ab initio} quantum chemistry package MOLPRO version 2006.1 \cite{Molpro2006}, as described in detail in \cite{Loreau2010a}. An adapted basis set consisting of the aug-cc-pv5Z basis set \cite{Dunning1989} supplemented by one contracted Gaussian function per orbital per atom up to $n=4$ has been used. The PEC have been calculated at the state-averaged complete active space self-consistent field (CASSCF) level and are shown in figure \ref{fig_PEC}.

This approach allows us to compute the non-adiabatic radial couplings $F_{mm^\prime}$, which are the matrix elements of the operator $\partial_R$ in the basis of the adiabatic electronic functions $\{\zeta_m\}$: $F_{mm^{\prime}}=\langle\zeta_m \vert \partial_{R} \vert \zeta_{m^{\prime}} \rangle$. The non-adiabatic radial couplings are used to build the diabatic representation \cite{Smith1969}, and the adiabatic-to-diabatic transformation matrix $\mathbb{D}(R)$ is the solution to the differential matrix equation
\begin{equation}
\partial_R\mathbb{D}(R)+\mathbb{F(R)\cdot D(R)}=0
\end{equation}
We have solved this equation starting from $R=\infty$, where we have the initial condition $\mathbb{D}(\infty)=\mathbb{I}$. The diabatic potential energy curves are the diagonal elements of the matrix $\mathbb{U}^\mathrm{d}=\mathbb{D}^{-1}\cdot \mathbb{U\cdot D}$, where $\mathbb{U}$ is the matrix of $H^{\mathrm{el}}$ in the adiabatic representation, while the off-diagonal elements of $\mathbb{U}^\mathrm{d}$ are the diabatic couplings. In the diabatic representation, the PEC cross and the couplings are a smooth function of the internuclear distance $R$.

In our description of the photodissociation process, we also included the non-adiabatic rotational couplings, which are the matrix elements of the operator $L_y$ and connect states with $\Delta \Lambda =1$. Finally, we computed the dipole matrix elements as they govern the photodissociation process (see below).

In Ref. \cite{Pedersen2010}, the authors underline the importance of states with $n>3$ during the photodissociation.
We checked this assumption by introducing the calculated adiabatic PEC for the $n=4$ $^1\Sigma^+$ and $^1\Pi$ states obtained in Ref. \cite{Loreau2010a} in our calculations. However, as we were unable to compute the non-adiabatic couplings between these states, we could only use these states to calculate the photodissociation cross section in the Born-Oppenheimer approximation (see section \ref{section_n=4}).

\subsection{Computation of the cross section}

The initial wave packet is constructed on each of the excited states by multiplying the wave function of the initial state $\psi_{vJ} (R)$, with energy $E_{vJ}$, by the matrix element $\mu_{10, m\Lambda^{\prime}}(R)$ of the dipole operator between the ground state ($m=1$, $\Lambda=0$) and an excited state $m$ of the $\Lambda^{\prime}$ symmetry,
\begin{equation} \label{photodiss_initial}
\Phi_{vJ}^{m\Lambda^{\prime}J^{\prime}}(R,t=0) = \mu_{10, m\Lambda^{\prime}}(R) \psi_{v}J(R) \ .
\end{equation}
The wave functions $\psi_{vJ} (R)$ and the energies $E_{vJ}$ were obtained by solving numerically the vibrational Schr\"odinger equation in the potential of the ground state using a B-spline basis set method as in Ref. \cite{Loreau2010c}.

The initial wave packet is propagated in time using the split operator \cite{Feit1982}, which requires the knowledge of the diabatic representation. The propagator is composed of the potential and kinetic energy operators, the action of which is carried out in the coordinate and momentum representations, respectively. The switch between the two representations is done using the fast Fourier transform algorithm. The time step was 0.2 a.u., or 4.84$\cdot 10^{-3}$ fs.
We used a spatial numerical grid covering distances from 0.5 to 100 a.u., divided into $2^{12}$ equally-spaced intervals.  In order to avoid reflections of the wave packet at the end of the grid, we introduced an absorbing potential in the asymptotic region at an internuclear distance $R_c=80$ a.u.

The total photodissociation cross section is obtained from the Fourier-transform of the autocorrelation function $C(t)= \sum_m \langle \Phi_{vJ}^{m\Lambda^{\prime}J^{\prime}}(R,0) \vert \Phi_{vJ}^{m\Lambda^{\prime}J^{\prime}}(R,t) \rangle$ \cite{Heller1978}. However, since the experiment distinguishes between the  dissociating channels He + H$^+$ or He$^+$ + H, we will only look at the partial cross sections, {\it i.e.} the contribution from each of the excited electronic states. These are obtained by a method using the Fourier transform of the asymptotic wave packets on the electronic states at an internuclear distance $R_{\infty}$, located in the asymptotic region but before the absorbing potential. At each time, the wave packet is analyzed at the point $R=R_\infty$ and the partial cross section as a function of the photon energy $E=h\nu$ for each excited state is given by \cite{BalintKurti1990}
\begin{equation}\label{photodiss_part_CS_R}
\sigma_{vJ\rightarrow m \Lambda^{\prime}J^{\prime}} (E) =
\frac{4\pi^2}{\mu \alpha a_0^2 k_m} E  \big\vert  A_{vJ}^{m \Lambda^{\prime}J^{\prime}}  \big\vert ^2
\end{equation}
where $\mu$ is the reduced mass, $k_m=\sqrt{2\mu(E-E^{\mathrm{as}}_m)}$ is the wave number in the electronic channel $m$ with an asymptotic energy $E^{\mathrm{as}}_m$, $\alpha$ is the fine structure constant, $a_0$ is the Bohr radius, and
\begin{equation}
A_{vJ}^{m \Lambda^{\prime}J^{\prime}} =
\frac{1}{\sqrt{2\pi}} \int_{0}^{\infty} \Phi_{vJ}^{m\Lambda^{\prime}J^{\prime}}(R_{\infty},t) e^{i(E_{vJ}+E)t} dt \ .
\end{equation}

Starting from an initial level $J$ in the ground state, the photoabsorption can excite the molecular ion into $^1\Sigma^+$ states with $J^{\prime}= J \pm 1$ or $^1\Pi$ states with $J^\prime = J \pm 0,1$. To obtain the photodissociation cross section for a given rotational level $J$, it is necessary to sum the various contributions as
 \begin{equation}\label{photo_tot_CS_J}
\sigma_{vJ \rightarrow m \Lambda^{\prime}} (E) = \sum_{J^{\prime}}\frac{S_{J\Lambda,J^{\prime}\Lambda^{\prime}}}{2J^{\prime}+1}  \sigma_{vJ \rightarrow m\Lambda^{\prime}J^{\prime}}(E)
\end{equation}
The H\"onl-London factors $S_{J\Lambda,J^{\prime}\Lambda^{\prime}}$ indicate how the intensity of a transition is distributed among the rotational branches and are given in Ref. \cite{Hansson2005}.

\section{Experimental determination of the ro-vibrational population of the ions}\label{vib_distrib}

\begin{figure}
\includegraphics[width=.4\textwidth]{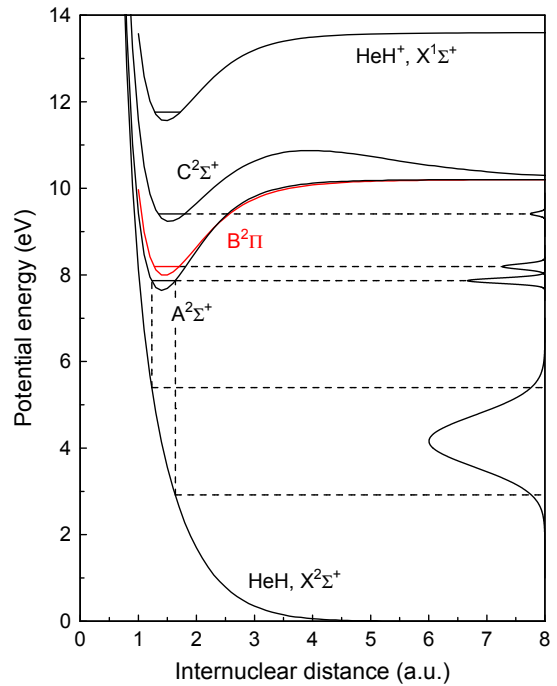}
\caption{Illustration of the decay of excited states of HeH following the charge transfer reaction (\ref{CX_HeH}) via photoemission or predissociation. The corresponding KER is shown on the right of the figure.}
\label{fig_heh}
\end{figure}

In order to determine the vibrational and rotational distribution prevailing under the conditions of the experiment of Pedersen {\it et al.}, we took advantage of the dissociative charge transfer set-up developed for the characterization of photoions \cite{Urbain04}. A keV beam of HeH$^+$ ions is accelerated out of a duoplasmatron source operating with a mixture of hydrogen and helium at similar pressure and arc current as in \cite{Pedersen2010}. The beam passes through an effusive potassium jet, where charge transfer takes place. As a result, excited HeH molecules are formed, that undergo predissociation towards the He + H ground state asymptote:
\begin{equation}\label{CX_HeH}
\mathrm{HeH}^+ + \mathrm{K} \longrightarrow (\mathrm{HeH})^* + \mathrm{K}^+ \longrightarrow \mathrm{He} + \mathrm{H} + \mathrm{K}^+
\end{equation}
The fragments are detected in coincidence by a pair of position sensitive detectors (Quantar), giving access to the total kinetic energy release (KER). This process has been extensively studied by van der Zande {\it et al.} \cite{vdZande}, who have identified the contribution of three different molecular states to the predissociation yield, namely the $A\ ^2\Sigma^+$, $C\ ^2\Sigma^+$, and $B\ ^2\Pi$ states. The mechanism is illustrated schematically in figure \ref{fig_heh}. Due to energy conservation, the kinetic energy release spectrum exhibits around 8 eV a group of peaks corresponding to the ro-vibrational ladder of an unknown combination of the  three states. In order to lift that major limitation of the method, we have set up an electrostatic ion beam trap consisting of a pair of mirror electrodes switched on and off at appropriate times to allow for injection and extraction of an ion bunch. With typical beam lifetimes of about 40 ms, we were able to let the ions cool down to the vibrational ground state before they undergo the charge transfer process in a separate chamber located downstream. After vibrational cooling, the  KER spectra are dominated by the predissociation of the $v=0$ level of the $A$, $B$ and $C$ molecular states, allowing for an accurate determination of their respective contribution to the signal, together with an effective rotational temperature.
The extracted weights, \textit{i.e.} $1:0.45:0.075$ for the $A$, $B$ and $C$ state respectively, are then used to fit a rotational and vibrational HeH$^+$ population to the hot spectra, making use of the Franck-Condon matrices connecting the ionic and neutral states. The choice of a Boltzmann distribution of rotational states, convolved with the experimental energy resolution, greatly simplifies the otherwise unstable fitting procedure.

The vibrational population extracted from measurements performed with a duoplasmatron source operating under conditions similar to the FLASH experiment \cite{Pedersen2010} is given in Table \ref{vib_distrib_table}. It is similar to the population obtained by van der Zande {\it et al.} \cite{vdZande} who assumed an equal contribution of the $A$ and $B$ states, namely a dominant population of $v=0$ with a monotonously decreasing population of higher states, limited to the first five vibrational levels. While this distribution is almost independent of the source parameters, an extensive rotational excitation is observed, which happens to be very sensitive to the type of ion source (ECR, duoplasmatron, hollow cathode) and its operating conditions, with temperatures ranging from 3000 to 5000 K.  Such a high temperature, already reported by several authors \cite{vdZande,Ketterle}, is not surprising since the HeH$^+$ creation process involves heavy particle collisions between vibrationally excited H$_2^+$ ions and He atoms. Conversely, some vibrational quenching may occur in high pressure sources, although no clear connection could be established between source conditions and internal energy of the ions. The value quoted here for a duoplasmatron source,  $T=3400 \pm 300$ K, falls in agreement with the temperature of 3100 K obtained by optical spectroscopy \cite{Ketterle}.

\begin{table}
\begin{center}
\begin{tabular}{cccc}
$\ \ v \ \ $	& Population		& $\sigma_{\mathrm{He} + \mathrm{H}^+}$ (cm$^2$)	& $\sigma_{\mathrm{He}^+ + \mathrm{H}}$ (cm$^2$)	 \\ \hline
0	& $0.55\pm 0.05$	& $2.90\cdot 10^{-18}$	& $1.23\cdot 10^{-18}$	\\ [0.3ex]
1	& $0.23\pm 0.05$	& $5.28\cdot 10^{-19}$ 	& $1.63\cdot 10^{-18}$	\\ [0.3ex]	
2	& $0.11\pm 0.05$	& $3.46\cdot 10^{-19}$ 	& $5.99\cdot 10^{-19}$	\\ [0.3ex]
3	& $0.07\pm 0.05$	& $3.25\cdot 10^{-19}$ 	& $3.49\cdot 10^{-19}$	\\ [0.3ex]
4	& $0.04\pm 0.05$	& $3.90\cdot 10^{-19}$ 	& $5.88\cdot 10^{-19}$	\\ [0.5ex] \hline\hline
\end{tabular}
\caption{Vibrational distribution of the ground $X\ ^1\Sigma^+$ state of HeH$^+$ for the experimental conditions of Ref. \cite{Pedersen2007} and theoretical cross sections for the two dissociating channels in (\ref{photodissociation}) at $h\nu =38.74$ eV. Error bars account for both the statistical error and the systematics introduced by the simultaneous fitting of a rotational temperature and a vibrational distribution. }
\label{vib_distrib_table}
\end{center}
\end{table}

\section{Results and discussion}	\label{results}

\subsection{Total cross section}

\begin{figure}
\centering
\includegraphics[angle=-90,width=.5\textwidth]{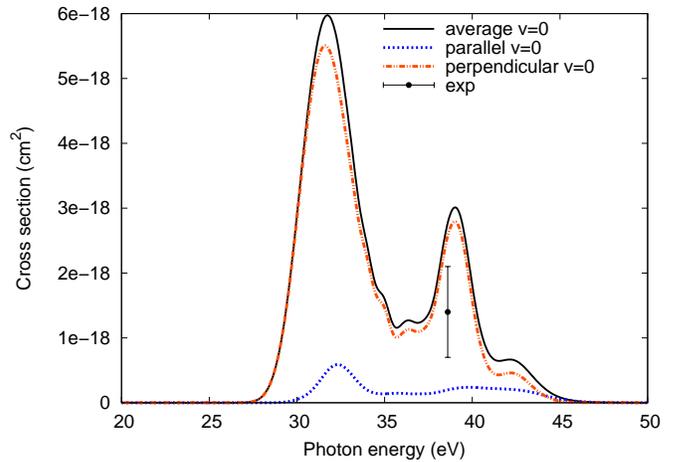}
\caption{Photodissociation cross section into He$(1snl\ ^1L)$ + H$^+$ for $v=0, J=0$.
The contribution from $^1\Sigma^+$ and $^1\Pi$ states (parallel and perpendicular orientation, respectively) is shown, as well as the cross section averaged over an isotropic orientation of the field and the experimental data with error bars.}
\label{CS_exp_v0}
\end{figure}

The total cross section for the photodissociation into He$(1snl\ ^1L)$ + H$^+$ starting from the initial state $v=0, J=0$ is shown in figure \ref{CS_exp_v0}, together with the experimental value, $\sigma = (1.4\pm 0.7)\cdot 10^{-18}$ cm$^2$. We see that the calculated cross section is larger than the upper limit of the experimental error bars. This result is slightly different than the one previously reported, the difference being due to an error in the energy scale  in Ref. \cite{Sodoga2009} and to the large variation of the cross section around the experimental energy (see figure \ref{CS_exp_v0}).

\begin{figure}
\centering
\includegraphics[angle=-90,width=.5\textwidth]{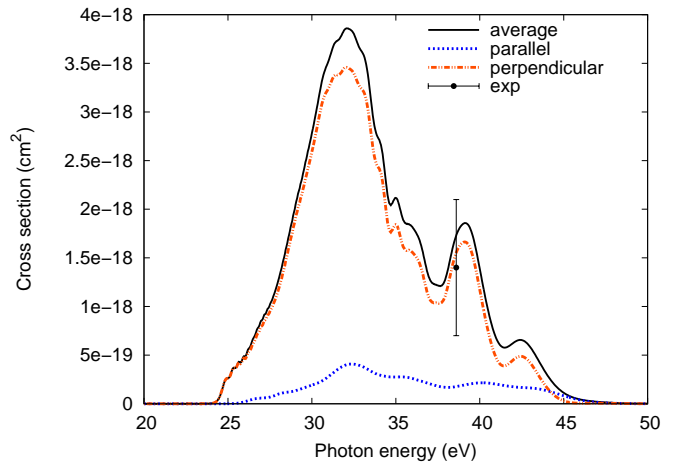}
\caption{Same as figure \ref{CS_exp_v0}, but for the vibrational distribution of table \ref{vib_distrib_table} and $J=0$. }
\label{CS_exp_tot}
\end{figure}

In addition to the vibrational distribution of the ions, we present in table \ref{vib_distrib_table} the cross sections for dissociation into He + H$^+$ or He$^+$ + H at the experimental energy.
The cross section for the photodissociation into He + H$^+$ weighted by the experimental vibrational distribution is presented in figure \ref{CS_exp_tot}. At 38.74 eV, we obtain a cross section of $\sigma = 1.78 \cdot 10^{-18}$ cm$^2$, close to the experimental value. At the experimental energy, while the $v=0$ cross section is larger than the experimental value, it is balanced by the fact that the cross section is much smaller for all the other values of $v$. An interesting observation is that at the experimental energy, the cross sections for $v>0$ all have very similar values (ranging from $3.25\cdot 10^{-19}$ to $5.28\cdot 10^{-19}$ cm$^2$), so that the main source of error in our theoretical estimation comes from the proportion of ions in $v=0$.

It should be noted that for $v=0$ or for the vibrational distribution of table \ref{vib_distrib_table}, we observe a dominance of the $^1\Pi$ states (perpendicular orientation) in the photodissociation process.

\subsection{Influence of the $n\geq 4$ states} \label{section_n=4}

One of the experimental observations is the important contribution arising from states with $n\geq3$ in the dissociation process, which was seen for both hot and cold ions. However, as indicated in Ref. \cite{Pedersen2010}, the origin of the dominant role played by these Rydberg states remains unclear.

In our calculations, we could only include states up to $n=3$, so that we cannot reproduce this experimental observation. This limitation of our approach is mainly due to the fact that we were unable to compute the non-adiabatic radial couplings between the $n>3$ states. However, as reported in \cite{Loreau2010a}, we were able to obtain the adiabatic PEC of the $n=4$ $^1\Sigma^+$ and $^1\Pi$ states, as well as their transition dipole moments with the ground state. With the PEC and dipole moments, we can calculate the photodissociation cross section in the Born-Oppenheimer approximation ({\it i.e.}, neglecting all non-adiabatic couplings). While the partial cross section will be incorrect (since the $n=4$ states are strongly coupled and that the couplings modify the final population of the electronic states), we can still study the effect of these states on the total cross section.

\begin{figure}
%\begin{minipage}[t]{1\textwidth}
%\centering
\includegraphics[angle= -90, width=.5\textwidth]{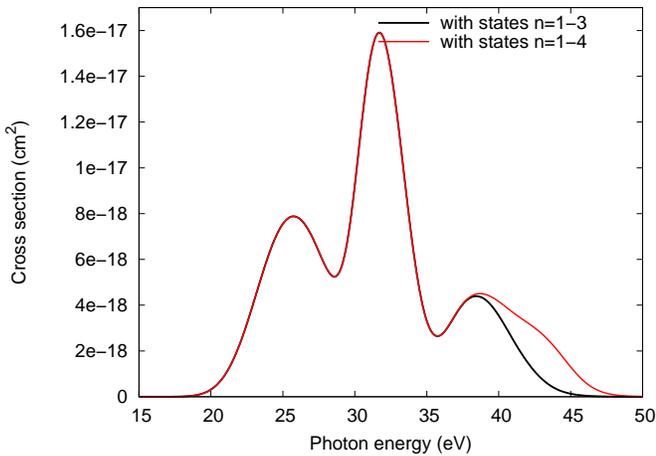}
\caption{Total photodissociation cross section for $v=0$, $J=0$ calculated in the Born-Oppenheimer approximation. Black line: using the $n=1-3$ sates; red line: using the $n=1-4$ states.}
\label{CS_comp_n_v0}
%\end{minipage}
%\begin{minipage}[t]{1\textwidth}
\end{figure}

\begin{figure}
%\vspace{0.5cm}
%\centering
\includegraphics[angle= -90, width=.5\textwidth]{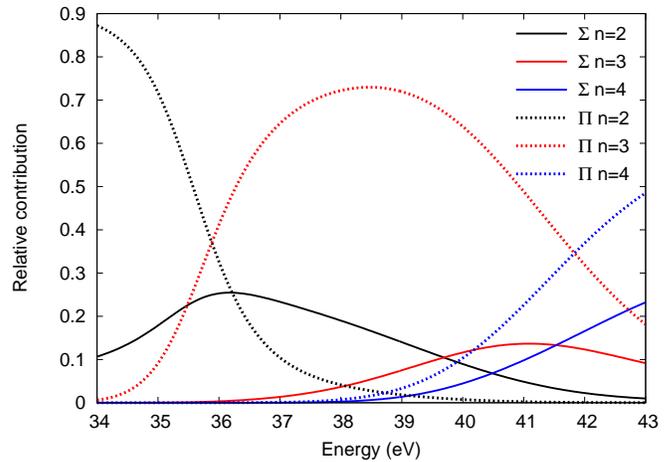}
\caption{Relative contribution of the $n=2-4$ manifolds to the total photodissociation cross section for $v=0$, as a function of the photon energy. The $n=1$ contribution is too small to be shown.}
\label{adia_n_v0}
%\end{minipage}
\end{figure}

The results are shown in figure \ref{CS_comp_n_v0} for $v=0$. We observe that the total cross section is modified by the inclusion of the $n=4$ states, but that it is virtually unaltered at the experimental energy and below. This can be understood by looking at the PEC presented in figure \ref{fig_PEC}: when a 32 nm photon is absorbed, the excitation occurs primarily in the $n=2-3$ states.
In figure \ref{adia_n_v0}, we show the relative contributions of the $^1\Sigma^+$ and $^1\Pi$ states to the cross section, grouped according to the value of the principal quantum number $n$. Around the experimental energy, the principal contribution is due to the $n=3$ $^1\Pi$ states, followed by the $n=2$ $^1\Sigma^+$ states.
From these figures, it is difficult to understand how states with $n\geq4$ could play such a major part in the photodissociation process for cold ions. A possible explanation would involve a population transfer from the $n=3$ to the $n>3$ states due to the non-adiabatic radial couplings. However, this explanation seems to be somewhat incompatible with the theoretical results published on the charge transfer in H($nl$) + He$^+(1s)$ collisions \cite{Loreau2010b}, which showed little interaction between manifolds of different value of $n$. % expand on that?
Moreover, by comparing the photodissociation cross section calculated in the Born-Oppenheimer approximation or with the non-adiabatic couplings included, we observed only a small transfer of population between the $n=2$ and $n=3$ states.


\begin{figure}
%\begin{minipage}[t]{1\textwidth}
%\centering
\includegraphics[angle= -90, width=.5\textwidth]{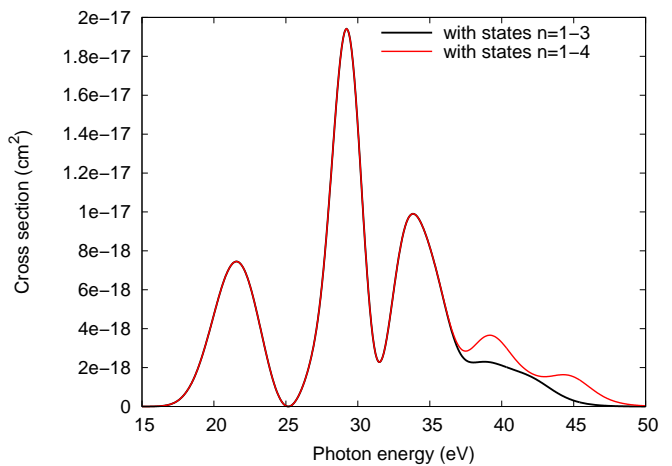}
\caption{Total photodissociation cross section for $v=1$, $J=0$ calculated in the Born-Oppenheimer approximation. Black line: using the $n=1-3$ sates; red line: using the $n=1-4$ states.}
\label{CS_comp_n_v1}
%\end{minipage}
%\begin{minipage}[t]{1\textwidth}
\end{figure}

\begin{figure}
%\vspace{0.5cm}
%\centering
\includegraphics[angle= -90, width=.5\textwidth]{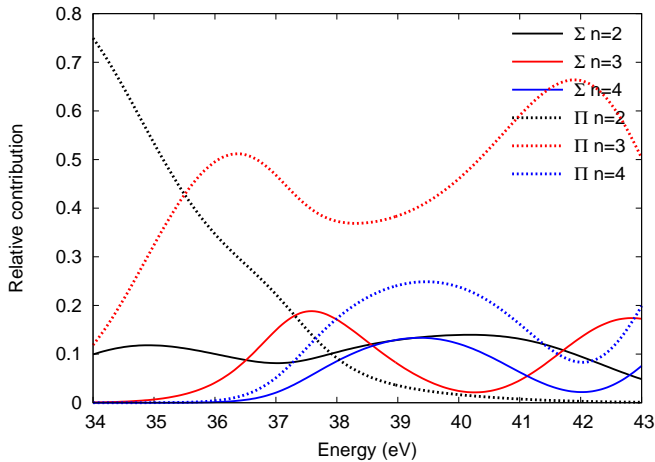}
\caption{Relative contribution of the $n=2-4$ manifolds to the total photodissociation cross section for $v=1$, as a function of the photon energy. The $n=1$ contribution is too small to be shown.}
\label{adia_n_v1}
%\end{minipage}
\end{figure}

Another striking result of the experimental study is the fact that the importance of highly excited states seems to be independent of the initial vibrational excitation of the molecular ion and of the dissociating channel, as can be seen on figures 6 and 7 of Ref. \cite{Pedersen2010}.
However, we expect that when the value of $v$ is increased, excited states with higher energy become accessible since the energy of the initial vibrational state, $E_{vJ}$, is larger. In addition, the vibrational wave functions for high $v$ cover a more important range of internuclear distances, so that the Franck-Condon region extends to higher value of $R$ and more excited states become directly accessible (see figure \ref{fig_PEC}).
To illustrate this phenomenon, we show in figures \ref{CS_comp_n_v1} and \ref{adia_n_v1} the contribution of the $n=4$ states to the cross section starting from the initial state $v=1$, calculated in the Born-Oppenheimer approximation. Compared to the $v=0$ case, we observe that at the experimental energy, the contribution of the $n=4$ states is more important. In particular, the $n=4$ $^1\Pi$ states account for about 25 \% of the total cross section. As a consequence, excited states with $n\geq 4$ should play a more important role in the photodissociation process for hot ions compared with cold ions, which is not observed experimentally.


\subsection{Branching ratios}

Pedersen {\it et al.} measured the branching ratio between the two dissociating channels in (\ref{photodissociation}), $\sigma_{\mathrm{He}^+ + \mathrm{H}}/\sigma_{\mathrm{He} + \mathrm{H}^+}$. They found a preference for dissociation into He$^+$ + H for cold ions, and an almost equal contribution from both channels for hot ions.
We calculated this branching ratio at the experimental energy for hot and cold ions by including the $n=1-3$ states using the values presented in table \ref{vib_distrib_table}. The results are presented in table \ref{branching_channels}.
We observe an important disagreement between theory and experiment, particularly for cold ions, which cannot be easily rationalized. For hot ions, the cross section for dissociation into the He + H$^+$ channel is in good agreement with the experiment (see figure \ref{CS_exp_tot}), which would point to an underestimate of the He$^+$ + H cross section. For cold ions, the experimental cross section is subject to a much larger uncertainty due to the smaller number of events detected. Unfortunately, the value of the cross section was not reported for these ions, so that a direct comparison with the experiment is not possible.
Moreover, as can be seen in figures \ref{CS_exp_v0} and \ref{CS_exp_tot}, the cross section for dissociation into He + H$^+$ is very sensitive to a shift in energy around the experimental value. As there is an spread in the photon energy of $\pm 0.6$ eV in FLASH, we checked the dependence of the results presented in table \ref{branching_channels} against such a shift. For example, if the photon energy is lowered by 0.6 eV, the ratio $\sigma_{\mathrm{He}^+ + \mathrm{H}}/\sigma_{\mathrm{He} + \mathrm{H}^+}$ is increased to 0.94 for cold ions and to 1.10 for hot ions.

\begin{table}
\begin{center}
\begin{tabular}{cccc}
			&  This work	& Experiment			 \\ \hline
Hot ions		& $0.66$		& $0.96 \pm 0.11$		 \\ [0.3ex]
Cold ions		& $0.42$		& $1.70 \pm 0.48$		 \\ [0.5ex] \hline\hline
\end{tabular}
\caption{Branching ratio $\sigma_{\mathrm{He}^+ + \mathrm{H}}/\sigma_{\mathrm{He} + \mathrm{H}^+}$ and comparison with Ref. \cite{Pedersen2010}.}
\label{branching_channels}
\end{center}
\end{table}

Pedersen {\it et al.} also estimated the contributions of $^1\Sigma^+$ and $^1\Pi$ states to the photodissociation process.
Our results are presented in table \ref{branching_angular_hot} for hot ions. We observe an excellent agreement with the experimental values for states dissociating into He$^+$ + H, but a quantitative disagreement for the He + H$^+$ channel. Qualitatively, we still confirm that the photodissociation in this channel is dominated by the $\Pi$ states.

However, the results of table \ref{branching_angular_hot} appear to contradict the above discussion. We found an good agreement with the experimental cross section for dissociation into the channel He + H$^+$, so that we would expect to obtain an agreement for the angular branching ratio as well. Indeed, the only mechanism allowing the transfer of population from $\Pi$ to $\Sigma$ states is via the non-adiabatic rotational couplings, but their effect was found to be negligible.
Moreover, to reproduce the experimental results of table \ref{branching_channels} for hot ions, we noted that it would be necessary to increase the cross section for dissociation into He$^+$ + H. However, the agreement between theory and experiment for the angular branching ratio is excellent. It would therefore be necessary to increase the cross section into He$^+$ + H in an equal part for $\Sigma$ and $\Pi$ states, which seems quite difficult to conceive.

\begin{table}
\begin{center}
\begin{tabular}{cccc}
			&			& This work	& Experiment	\\ [0.5ex] \hline
He + H$^+$	& $\Sigma$	& 10 \%	& $30 \pm 2$ \%	\\ [0.5ex]
			& $\Pi$		& 90 \%	& $70 \pm 2$ \%	\\ [0.5ex] \hline
He$^+$ + H	& $\Sigma$	& 39 \%	& $38 \pm 3$ \%	\\ [0.5ex]
			& $\Pi$		& 61 \% 	& $62 \pm 1$ \%	\\ [0.7ex] \hline\hline
\end{tabular}
\caption{Contribution of $^1\Sigma^+$ and $^1\Pi$ states to the photodissociation cross section for vibrationally hot ions. Comparison with the experimental values of \cite{Pedersen2010} for $J=0$. }
\label{branching_angular_hot}
\end{center}
\end{table}

The angular branching ratios for cold ions are presented in table \ref{branching_angular_cold}. As in the case of hot ions, the agreement is excellent for the He$^+$ + H channel but it is only qualitative for the He + H$^+$ channel. In the latter case, we confirm the dominant contribution from the $\Pi$ states, but this contribution is found to be larger than in the experiment.

Until now, we have assumed the effects of rotational excitation to be negligible, as found in Refs. \cite{Dumitriu2009} and \cite{Sodoga2009}. However, due to the high rotational temperature of the ions (see section \ref{vib_distrib}), we decided to investigate its role in the photodissociation process for cold ions into detail. Even though the cross section does not depend strongly on $J$, an effect is expected for large $J$ due to the change in the energy of the initial state.

If we assume a Maxwell-Boltzmann distribution, the population $p_J(T)$ of the rotational level $J$ with energy $E_J$ is given by
\begin{equation}\label{Max_Boltz}
p_J(T) = \frac{(2J+1)e^{-E_J/kT}}{\sum_{J{^\prime}} (2J{^\prime}+1)e^{-E{_J{^\prime}}/kT}}
\end{equation}
where $k$ is the Boltzmann constant. As indicated in section \ref{vib_distrib}, the temperature is $T\sim3400$ K.
The cross section for a Maxwell-Boltzmann distribution of initial rotational states is then given as
\begin{equation}\label{sigma_MB}
\overline{\sigma}=\sum_{J=0}^{J_{\max}} p_J(T) \sigma_J
\end{equation}
We computed the photodissociation cross sections $\sigma_J$ for excited rotational states up to $J_{\max}=22$.
The angular branching ratios for the weighted cross section $\overline{\sigma}$ are presented in table \ref{branching_angular_cold}. We observe that the effect is indeed very small. Moreover, this is also the case for the total cross section and the ratio $\sigma_{\mathrm{He}^+ + \mathrm{H}}/\sigma_{\mathrm{He} + \mathrm{H}^+}$.

As $J$ increases, the effect of the rotational couplings can also be expected to increase. This would modify the angular distribution, as the action of these couplings is to transfer population from $\Sigma$ to $\Pi$ states (and vice versa). However, we observed that even for large values of $J$, the rotational couplings do not modify significantly the cross sections, so that the discrepancies between experiment and theory do not appear to be related to a mechanism involving these couplings.

\begin{table}
\begin{center}
\begin{tabular}{ccccc}
			&			& This work, $J$=0	& This work, $p_J$	& Exp.	 \\ [0.5ex] \hline
He + H$^+$	& $\Sigma$	& 7 \%	& 8 \% 	& $24 \pm 6$ \%	\\ [0.5ex]
			& $\Pi$		& 93 \%	& 92 \% 	& $76 \pm 6$ \%      	\\ [0.5ex] \hline
He$^+$ + H	& $\Sigma$	& 46 \%	& 50 \% 	& $50 \pm 3$ \%	\\ [0.5ex]
			& $\Pi$		& 54 \%	& 50 \% 	& $50 \pm 5$ \% 	\\ [0.7ex] \hline\hline
\end{tabular}
\caption{Contribution of $^1\Sigma^+$ and $^1\Pi$ states to the photodissociation cross section for vibrationally cold ($v=0$) ions and comparison with the experimental values of Ref. \cite{Pedersen2010}. The results presented in the first column are for an initial rotational number $J=0$, while those in the  second column are for the Maxwell-Boltzmann distribution (\ref{Max_Boltz}).}
\label{branching_angular_cold}
\end{center}
\end{table}

Finally, we checked the dependence of the angular branching ratio against a shift in energy of about 0.6 eV, corresponding to the uncertainty on the FLASH photon energy, as we noted (see above) that this could improve the ratio of the cross sections $\sigma_{\mathrm{He}^+ + \mathrm{H}}/\sigma_{\mathrm{He} + \mathrm{H}^+}$. While the values presented in tables \ref{branching_channels} - \ref{branching_angular_cold} are modified by such a shift, this does not resolve the discrepancies between our calculations and the experimental values.

In Ref. \cite{Pedersen2010}, the experimental results are compared with the theoretical results of Dumitriu and Saenz \cite{Dumitriu2009} and an excellent agreement for the ratio  $\sigma_{\mathrm{He}^+ + \mathrm{H}}/\sigma_{\mathrm{He} + \mathrm{H}^+}$ is found for vibrationally cold ions.
As pointed out by Pedersen {\it et al.}, in the work of Dumitriu and Saenz there is a clear separation in energy between states dissociating into He + H$^+$ or He$^+$ + H (cf. figure 1 of Ref. \cite{Dumitriu2009}). Therefore, these authors conclude that the non-adiabatic couplings will not affect the ratio between the fragments and only take into account a single non-adiabatic radial coupling between two $n=2$ states. However, it was shown by Green {\it et al.} \cite{Green1974} that the first part of the electronic spectrum up to $n=3$ results from an alternation of states dissociating into He + H$^+$ or He$^+$ + H. This was confirmed and extended to $n=4$ states in the recent study of Loreau {\it et al.} \cite{Loreau2010a}. Furthermore, the electronic states are strongly coupled, so that the non-adiabatic couplings modify substantially the partial photodissociation cross sections.

\section{Summary and conclusions}

In this paper, we have expanded on a previous theoretical investigation of the photodissociation of HeH$^+$ in order to establish a detailed comparison with recent experiments \cite{Pedersen2007,Pedersen2010} for vibrational hot or cold ions. We calculated the photodissociation cross section by propagating wave packets on the coupled potential energy curves of the excited states of the ion. In order to compare our theoretical results with the experimental data, we have measured the ro-vibrational distribution of the ions using the same source conditions as in the experiment.

With the vibrational distribution, we have obtained a cross section for dissociation into the He + H$^+$ channel close to the reported experimental value. To get a good agreement between theory and experiment, it was necessary to take the non-adiabatic radial and rotational couplings into account.
On the other hand, our value for the cross section for dissociation into the other channel, He$^+$ + H, is smaller than the experimental one. For cold ions, only the ratio of the cross sections for the two dissociative channels, $\sigma_{\mathrm{He}^+ + \mathrm{H}}/\sigma_{\mathrm{He} + \mathrm{H}^+}$, has been reported. Our calculated value is in disagreement with the experimental estimate. To further compare the theory and experiment, it would be necessary to know the absolute value of the cross sections.
For each dissociative channel, we have also calculated the angular branching ratios. For both hot and cold ions, the agreement is excellent for dissociation into the He$^+$ + H channel but it is only qualitative for the He + H$^+$ channel. This finding is difficult to explain since the total cross section for He$^+$ + H is too small compared to the experimental value, while it is in good agreement for He + H$^+$. The only way to modify the angular branching ratios while leaving the total cross section unchanged involves the non-adiabatic rotational couplings; however, their effect was found to be negligible even for large $J$. Moreover, the effect of rotational excitation was found to be very small despite the high rotational temperature of the ions in the source.

In our opinion, the most surprising conclusion of the experiment is the importance of states with a principal quantum number $n>3$ both for hot and cold ions. While we were limited to the $n=1-3$ states due to the difficulty of calculating the non-adiabatic radial couplings for higher lying states, we could still compute the photodissociation cross section including the $n=4$ states in the Born-Oppenheimer approximation (that is, neglecting the radial couplings).
As shown in figures \ref{CS_comp_n_v0} and \ref{adia_n_v0}, the contribution of the $n=4$ states is negligible for $v=0$ at the experimental energy. Moreover, we argued that the importance of the $n>3$ states should increase with the vibrational excitation of the ions, something that is not seen in the experiment.
The excess kinetic energy associated with higher lying initial vibrational levels ($\approx 0.35$ eV per level) and higher rotational temperature could easily mask the downward shift of the dissociation energy with increasing principal quantum number of the products. These counteracting effects require further modeling, which lies beyond the scope of the present paper. This therefore remains an open question.

Due to the complexity of the experiment, there is however some uncertainty on the results which could affect the comparison. As noted by Pedersen {\it et al.}, direct ionization of HeH$^+$ into He$^+(1s)$ + H$^+$ becomes possible at the experimental energy for vibrationally excited ions. This would lead to events appearing as H$^+$, which would modify the branching ratios. However, this would not resolve the discrepancies observed for vibrationally  cold ions.
Another effect that would affect the determination of the branching ratios is the possible presence of the $a\ ^3\Sigma^+$ state of HeH$^+$ in the source.
As it has a very long lifetime \cite{Loreau2010c}, this state (which lies about 11 eV above the ground state) would be present for both hot and cold ions. The absorption of a 38.7 eV photon by ions in this electronic state would lead to direct ionization, and thus to the observation of He$^+(1s)$ + H$^+$. The $a\ ^3\Sigma^+$ state has been previously observed in HeH$^+$ sources \cite{Strasser2000}. Its presence was inferred from the low kinetic energy released by dissociative charge transfer with an Ar target, and its disappearance with ion storage time. We have repeated this experiment and replaced Ar with He, making the charge transfer process more resonant for the $a\ ^3\Sigma^+$ state of HeH$^+$ than for its ground state.  No sizeable effect was observed, which contradicts the attribution of the low KER to the $a\ ^3\Sigma^+$ state.  It seems thus unlikely that HeH$^+$ is present in the $a\ ^3\Sigma^+$ in an amount that would significantly modify the above results.

In conclusion, while we have been able to explain some of the experimental findings, there remains discrepancies between theory and experiment. Despite the simplicity of the system considered, the fact that a large number of excited electronic states play a part in the photodissociation process makes an accurate comparison difficult to achieve.
From an experimental perspective, it would be very interesting to measure the cross sections and branching ratios at lower energy in order to make the comparison with theory easier. In particular, we observed that the cross section changes very quickly in the region around the experimental photon energy.
From a theoretical point of view, the problem is complicated by the fact that the non-adiabatic radial couplings cannot be neglected in the treatment of the photodissociation. To investigate the role of the $n>3$ states, it would therefore be necessary to calculate explicitly these couplings.

\acknowledgments

We would like to thank H.B. Pedersen for providing us the conditions of operation of the ion source used in the FLASH experiment. We are also grateful to M. Desouter-Lecomte, D. Lauvergnat and Y. Justum for their help with the computer codes.
We acknowledge financial support from the F.R.S.-FNRS under I.I.S.N. contract n$^\circ$ 4.4504.10 and of the ARC (ATMOS) and FRFC programs.
J. Loreau is supported by a fellowship from the Belgian American Educational Foundation. X.U. is Senior Research Associate of the Fund for Scientific Research - FNRS.

%\clearpage


\begin{thebibliography}{99}


\bibitem{Lepp2002} S. Lepp, P. C. Stancil, and A. Dalgarno, {\it J. Phys. B: At. Mol. Opt. Phys.} {\bf 35}, R57 (2002).
\bibitem{Dabrowski1978} I. Dabrowski and G. Herzberg, {\it Trans. NY Acad. Sci.} {\bf 38}, 14 (1978).
\bibitem{Harris2004} G. J. Harris, A. E. Lynas-Gray, S. Miller, and J. Tennyson, {\it Ap. J.} {\bf 617}, L143 (2004).
\bibitem{Engel2005} E. Engel, N. Doss, G. J. Harris, and J. Tennyson, {\it Mon. Not. R. Astron. Soc.} {\bf 357}, 471 (2005).
\bibitem{Liu1997} X.-W. Liu, M.J. Barlow, A. Dalgarno, J. Tennyson, T. Lim, B.M. Swinyard, J. Cernicharo, P. Cox, J.-P. Baluteau, D. P\'equignot, Nguyen-Q-Rieu, R.J. Emery and P.E. Clegg, {\it Mon. Not. R. Astron. Soc. \bf 290} L71 (1997).
\bibitem{Saha1978} S. Saha, K. K. Datta, and A. K. Barua, {\it J. Phys. B: At. Mol. Phys.} {\bf 11}, 3349 (1978).
\bibitem{Roberge1982} W. Roberge and A. Dalgarno, {\it Ap. J.} {\bf 255}, 489 (1982).
\bibitem{Basu1984} D. Basu and A. K. Barua, {\it J. Phys. B: At. Mol. Phys.} {\bf 17}, 1537 (1984).
\bibitem{Pedersen2007} H. B. Pedersen, S. Altevogt, B. Jordon-Thaden, O. Heber, M.L. Rappaport, D. Schwalm, J. Ullrich, D. Zajfman, R. Treusch, N. Guerassimova, N. Martins, J.-T. Hoeft, M. Wellhofer, and A. Wolf, {\it Phys. Rev. Lett.} {\bf 98}, 223202 (2007).
\bibitem{Dumitriu2009} I. Dumitriu and A. Saenz, {\it J. Phys. B: At. Mol. Opt. Phys.} {\bf 42}, 165101 (2009).
\bibitem{Sodoga2009} K. Sodoga, J. Loreau, D. Lauvergnat, Y. Justum, N. Vaeck, and M. Desouter-Lecomte, {\it Phys. Rev. A} {\bf 80}, 033417 (2009).
\bibitem{Pedersen2010} H. B. Pedersen, L. Lammich, C. Domesle, B. Jordon-Thaden, O. Heber, J. Ullrich, R. Treusch, N. Guerassimova and A. Wolf, {\it Phys. Rev. A} {\bf 82}, 023415 (2010).
\bibitem{Loreau2010a} J. Loreau, P. Palmeri, P. Quinet, J. Li\'evin, and N. Vaeck, {\it J. Phys. B: At. Mol. Opt. Phys.} {\bf 43}, 065101 (2010).
\bibitem{Molpro2006} H.-J. Werner, P.J. Knowles, R. Lindh, F.R. Manby, M. Sch\"utz, and others. MOLPRO, version 2006.1, a package of {\it ab initio} programs, see http://www.molpro.net
\bibitem{Dunning1989} T. H. Dunning, Jr., {\it J. Chem. Phys.} {\bf 90}, 1007 (1989).
\bibitem{Smith1969} F. T. Smith, {\it Phys. Rev.} {\bf 179}, 111 (1969).
\bibitem{Loreau2010c} J. Loreau, J. Li\'evin, and N. Vaeck, {\it J. Chem. Phys.} {\bf 133} 114302 (2010).
\bibitem{Feit1982} M. D. Feit, J. A. Fleck Jr., and A. Steiger, {\it J. Comput. Phys.} {\bf 47}, 412 (1982).
\bibitem{Heller1978} E. Heller, {\it J. Chem. Phys.} {\bf 68}, 2066 (1978).
\bibitem{BalintKurti1990} G. G. Balint-Kurti, R. N. Dixon, and C. Marston, {\it J. Chem. Soc. Faraday Trans.} {\bf 86}, 1741 (1990).
\bibitem{Hansson2005} A. Hansson and J. Watson, {\it J. Mol. Spectrosc.} {\bf 233}, 169 (2005).
\bibitem{Urbain04} X. Urbain, B. Fabre, E. M. Staicu-Casagrande, N. de Ruette, V. M. Andrianarijaona, J. Jureta, J. H. Posthumus, A. Saenz, E. Baldit and C. Cornaggia, {\em Phys. Rev. Lett.} {{\bf 92} 163004} (2004)
\bibitem{vdZande} W. J. van der Zande, W. Koot, D.P. de Bruijn, and C. Kubach, {\em Phys. Rev. Lett.} {{\bf 57} 1219} (1986).
\bibitem{Ketterle} W. Ketterle, H. Figger, and H. Walther {\em Phys. Rev. Lett.} {{\bf 55} 2941} (1985).
\bibitem{Loreau2010b} J. Loreau, K. Sodoga, D. Lauvergnat, M. Desouter-Lecomte, and N. Vaeck, {\it Phys. Rev. A} {\bf 82}, 012708 (2010).
\bibitem{Green1974} T. A. Green, J. C. Browne, H. H. Michels, and M. M. Madsen, {\it J. Chem. Phys.} {\bf 61}, 5186 (1974);
{\it J. Chem. Phys.} {\bf 61}, 5198 (1974); T. A. Green, H. H. Michels, and J. C. Browne, {\it J. Chem. Phys.} {\bf 64}, 3951 (1976); {\it J. Chem. Phys.} {\bf 69}, 101 (1978).
\bibitem{Strasser2000} D. Strasser, K.G. Bhushan, H.B. Pedersen, R. Wester, O. Heber, A. Lafosse, M.L. Rappaport, N. Altstein, and D. Zajfman, {\it Phys. Rev. A \bf 61} 060705 (2000)

\end{thebibliography}
\end{document}